\documentclass{elsart}
\usepackage{cite}
\usepackage{graphicx}
\usepackage{dcolumn}
\usepackage{amsmath}

\begin{document}

\begin{frontmatter}

\title{On the Hellmann-Feynman theorem in statistical mechanics}

\author{Paolo Amore \thanksref{PA}}

\address{Facultad de Ciencias, CUICBAS, Universidad de Colima,Bernal D\'{\i}az del Castillo 340, Colima,
Colima,Mexico}

\author{Francisco M. Fern\'andez \thanksref{FMF}}

\address{INIFTA, Divisi\'{o}n Qu\'{\i}mica Te\'{o}rica,\\
Blvd. 113 y 64 (S/N), Sucursal 4, Casilla de Correo 16,\\
1900 La Plata, Argentina}

\thanks[PA]{e--mail: paolo@ucol.mx}
\thanks[FMF]{e--mail: fernande@quimica.unlp.edu.ar}

\begin{abstract}
In this paper we develop the Hellmann-Feynman theorem in
statistical mechanics without resorting to the eigenvalues and
eigenvectors of the Hamiltonian operator. Present approach does
not require the quantum-mechanical version of the theorem at $T=0$
and bypasses any discussion about degenerate states.
\end{abstract}


\end{frontmatter}

\section{Introduction}

\label{sec:intro}

Many years ago Feynman\cite{F39} developed a method for the calculation of
forces in molecules that does not require the explicit use of the derivative
of the energy. This expression, known as the Hellmann-Feynman theorem (HFT),
is discussed in almost every book on quantum mechanics and quantum chemistry%
\cite{CDL77,P68} and some pedagogical articles discuss its utility
in quantum mechanics\cite{E54,BS89}. The HFT has also been applied
to perturbation theory\cite{E54}, even for degenerate
states\cite{BS89}. It is worth pointing out that the theorem had
been developed by G\"{u}ttinger\cite {G32} several years earlier
(and in its more general off-diagonal form!!). However, in what
follows we adhere to the common usage and will refer to the
theorem by the usual and widely accepted name.

Some time ago Zhang and George\cite{ZG02} reported a supposedly failure of
the theorem in the case of degenerate states and proposed a remedy. Such
assessment resulted curious in the light that the proof of the theorem does
not require that the states are nondegenerate\cite{F39,CDL77,P68,E54,BS89}.
Several authors commented on this paper proving Zhang and George wrong with
respect to the failure of the HFT\cite{AC03,F04,V04,BHM04}. In particular,
Fern\'{a}ndez\cite{F04} showed that the expression for the supposed remedy
is correct but unnecessary because the original diagonal HFT is valid for
degenerate states provided that one chooses the correct linear combinations
of the degenerate eigenfunctions for the calculation. Such linear
combinations appear naturally when one considers the off-diagonal version of
the HFT and the limit $\lambda \rightarrow \lambda _{0}$, where $\lambda $
is a parameter in the Hamiltonian operator $H$ and $\lambda _{0}$ the
particular value of $\lambda $ at which degeneracy occurs\cite{F04}. Despite
all these proofs of the validity of the HFT the problem still seems to be
poorly understood\cite{R07,RS19}. This fact motivated a recent article with
the purpose of clarifying the main points about the application of the HFT
to degenerate states\cite{F19}.

There has also been interest in the form of the HFT in statistical mechanics
where the authors have resorted to suitable sums over eigenstates of the
Hamiltonian operator in their calculations\cite{FC95,R07,F14b} (and
references therein). In particular, Rai\cite{R07} took into account the
degenerate subspaces explicitly and argued that his results were a
generalization of those of Fan and Chen\cite{FC95} who had not considered
degeneracy explicitly. More precisely, the former author resorted to the
remedy of Zhang and George\cite{ZG02} that, as has already been proved, is
unnecessary if one assumes that the mathematical procedure of Fan and Chen%
\cite{FC95} is based on the degenerate eigenstates of the Hamiltonian
operator that satisfy the HFT\cite{F04}. It is worth noticing that both
strategies made use of the quantum-mechanical HFT ($T=0$) in their
derivation.

The purpose of this paper is to show that one can easily obtain
those results without using the eigenvalues and eigenvectors of
the Hamiltonian operator and therefore bypass the
quantum-mechanical HFT. In section\-~\ref {sec:HFT statistical} we
develop the HFT in the canonical ensemble by means of an arbitrary
basis set of states that do not depend on the parameter that is
commonly varied to derive the HFT in quantum and statistical
mechanics. One of the results in this section enables one to
derive the HFT in the grand canonical ensemble. Finally, in
section~\ref{sec:conclusions} we summarize the main results and
draw conclusions.

\section{The Hellmann-Feynman theorem for the canonical ensemble}

\label{sec:HFT statistical}

The statistical average of an observable $A$ in the canonical ensemble is
given by
\begin{equation}
\left\langle A\right\rangle _{av}=\mathrm{tr}\left( \rho A\right) ,\;\rho =%
\frac{e^{-\beta H}}{Z},\;Z=\mathrm{tr}\left( e^{-\beta H}\right) ,
\label{eq:<A>}
\end{equation}
where $H$ is the Hamiltonian operator of the system and $\beta =1/(k_{B}T)$.

Suppose that a function $f(H)$ can be expanded as
\begin{equation}
f(H)=\sum_{k=0}^{\infty }f_{k}H^{k},  \label{eq:f_expansion}
\end{equation}
and that $H$ depends on a parameter $\lambda $. Therefore,
\begin{equation}
\frac{\partial }{\partial \lambda }f(H)=\sum_{k=0}^{\infty
}f_{k}\sum_{i=0}^{k-1}H^{i}H^{\prime }H^{k-i-1},  \label{eq:f'}
\end{equation}
where the prime stands for derivative with respect to $\lambda $. Obviously,
we are assuming that the coefficients $f_{j}$ do not depend on $\lambda $.

Taking into account a well known property of the trace
\begin{equation}
\mathrm{tr}\left( H^{i}H^{\prime }H^{k-i-1}\right) =\mathrm{tr}\left(
H^{\prime }H^{k-1}\right) ,  \label{eq:tr(HHH)}
\end{equation}
we conclude that
\begin{equation}
\mathrm{tr}\left[ \frac{\partial }{\partial \lambda }f(H)\right] =\mathrm{tr}%
\left[ H^{\prime }D_{H}f(H)\right] ,  \label{eq:tr(f')}
\end{equation}
where
\begin{equation}
D_{H}f(H)=\left. \frac{\partial f(x)}{\partial x}\right| _{x=H}.
\label{eq:f'_def}
\end{equation}
It is clear that equation (\ref{eq:tr(f')}) is valid even when $[H^{\prime
},H]\neq 0$. Since the basis set used in the calculation of the trace is
arbitrary we assume that it is independent of $\lambda $ and equation (\ref
{eq:tr(f')}) becomes
\begin{equation}
\frac{\partial }{\partial \lambda }\mathrm{tr}\left[ f(H)\right] =\mathrm{tr}%
\left[ H^{\prime }D_{H}f(H)\right] ,  \label{eq:tr(f)'}
\end{equation}
that is the main result of this paper which will enable us to derive all the
other necessary expressions. For example,
\begin{equation}
\frac{\partial }{\partial \lambda }Z=Z^{\prime }=-\beta \mathrm{tr}\left(
H^{\prime }e^{-\beta H}\right) ,  \label{eq:Z'}
\end{equation}
and
\begin{eqnarray}
\frac{\partial }{\partial \lambda }\left\langle H\right\rangle _{av} &=&-%
\frac{Z^{\prime }}{Z^{2}}\mathrm{tr}\left( He^{-\beta H}\right) +\frac{1}{Z}%
\mathrm{tr}\left( H^{\prime }e^{-\beta H}\right) -\frac{\beta }{Z}\mathrm{tr}%
\left( HH^{\prime }e^{-\beta H}\right)  \nonumber \\
&=&\left\langle H^{\prime }\right\rangle _{av}+\beta \left( \left\langle
H\right\rangle _{av}\left\langle H^{\prime }\right\rangle _{av}-\left\langle
HH^{\prime }\right\rangle _{av}\right) .  \label{eq:<H>'}
\end{eqnarray}
Since we have derived the result of Fan and Chen\cite{FC95} and
Rai\cite{R07} without resorting to the eigenstates of $H$, it is
clear that it is valid whether there are degenerate states or not.
Equation (\ref{eq:tr(HHH)}) is strictly valid in the case of a
space of finite dimension. If the dimension is infinite we may
proceed as indicated in the Appendix~\ref{sec:appendix}.

The result above can be simplified a little bit further by means of the
relationship
\begin{equation}
\frac{\partial }{\partial \beta }\left\langle H^{\prime }\right\rangle
_{av}=\left\langle H\right\rangle _{av}\left\langle H^{\prime }\right\rangle
_{av}-\left\langle HH^{\prime }\right\rangle _{av},
\end{equation}
so that
\begin{equation}
\frac{\partial }{\partial \lambda }\left\langle H\right\rangle
_{av}=\left\langle H^{\prime }\right\rangle _{av}+\beta \frac{\partial }{%
\partial \beta }\left\langle H^{\prime }\right\rangle _{av}.
\label{eq:<H>'_b}
\end{equation}

In what follows we discuss the variation of the statistical
average of an observable $A$ with respect to $\lambda $ although
it has nothing to do with the HFT. One of the reasons for the
analysis of this problem is that it was discussed by Fan and
Chen\cite{FC95}, the other is that the result will prove useful
for developing the form of the HFT in the grand canonical
ensembel. For simplicity, we first consider an observable $A$ that
commutes with $H$ ($[H,A]=0$). On arguing as before we have
\begin{equation}
\mathrm{tr}\left( H^{i}H^{\prime }H^{k-i-1}A\right) =\mathrm{tr}\left(
H^{\prime }H^{k-i-1}AH^{i}\right) =\mathrm{tr}\left( H^{\prime
}AH^{k-1}\right) ,
\end{equation}
so that
\begin{equation}
\frac{\partial }{\partial \lambda }\mathrm{tr}\left[ f(H)A\right] =\mathrm{tr%
}\left[ H^{\prime }AD_{H}f(H)\right] .
\end{equation}
Therefore,
\begin{eqnarray}
\frac{\partial }{\partial \lambda }\left\langle A\right\rangle _{av} &=&-%
\frac{Z^{\prime }}{Z^{2}}\mathrm{tr}\left( e^{-\beta H}A\right) -\frac{\beta
}{Z}\mathrm{tr}\left( H^{\prime }Ae^{-\beta H}\right) +\frac{1}{Z}\mathrm{tr}%
\left( e^{-\beta H}A^{\prime }\right)  \nonumber \\
&=&\left\langle A^{\prime }\right\rangle _{av}+\beta \left( \left\langle
A\right\rangle _{av}\left\langle H^{\prime }\right\rangle _{av}-\left\langle
AH^{\prime }\right\rangle _{av}\right) .  \label{eq:<A>'}
\end{eqnarray}
This expression is a particular case of the one developed by Fan and Chen%
\cite{FC95} and both agree when the third and fourth terms in their equation
(15) vanish.

If $[H,A]\neq 0$ we can obtain a compact expression by defining the operator
$F$ as $\frac{\partial }{\partial \lambda }e^{\beta H}=Fe^{\beta H}$.
Differentiating both sides of $e^{\beta H}e^{-\beta H}=1$ with respect to $%
\lambda $ we obtain $\frac{\partial }{\partial \lambda }e^{-\beta
H}=-e^{-\beta H}F$ so that the derivative of $\left\langle A\right\rangle
_{av}$ becomes
\begin{equation}
\frac{\partial }{\partial \lambda }\left\langle A\right\rangle
_{av}=\left\langle A^{\prime }\right\rangle _{av}+\beta \left\langle
A\right\rangle _{av}\left\langle H^{\prime }\right\rangle _{av}-\left\langle
FA\right\rangle _{av}.
\end{equation}

Equation (\ref{eq:<A>'}) enables one to derive the HFT in the grand
canonical ensemble. To this end we turn to the notation of Rai\cite{R07} and
write
\begin{equation}
\left[ H\right] _{G}=\mathrm{tr}\left( \rho H\right) ,\;\rho =\frac{%
e^{-\beta K}}{Z_{G}},\;K=H-\mu \hat{N},\;Z_{G}=\mathrm{tr}\left( e^{-\beta
K}\right) ,
\end{equation}
where $H$ is the Hamiltonian operator in the Fock space, $\mu $ is the
chemical potential and $\hat{N}$ a particle-number operator. It is clear
that $\left[ H,\hat{N}\right] =0$ because these operators have a common set
of eigenvectors and $\partial \hat{N}/\partial \lambda =0$. Therefore, $%
\left[ K,H\right] =0$ and $K^{\prime }=H^{\prime }$ so that we can
substitute $K$ and $H$ for $H$ and $A$, respectively, in equation (\ref
{eq:<A>'}) thus obtaining Rai's result\cite{R07}
\begin{equation}
\frac{\partial }{\partial \lambda }\left[ H\right] _{G}=\left[ H^{\prime
}\right] _{G}+\beta \left( \left[ H\right] _{G}\left[ H^{\prime }\right]
_{G}-\left[ HH^{\prime }\right] _{G}\right) .
\end{equation}

In order to test equation (\ref{eq:<H>'_b}) we consider the dimensionless
harmonic oscillator
\begin{equation}
H=-\frac{1}{2}\frac{d^{2}}{dx^{2}}+\frac{1+\lambda }{2}x^{2}.
\end{equation}
In this case we have
\begin{equation}
\left\langle H\right\rangle _{av}=\frac{\sqrt{\lambda +1}\left( e^{\beta
\sqrt{\lambda +1}}+1\right) }{2\left( e^{\beta \sqrt{\lambda +1}}-1\right) },
\end{equation}
and
\begin{equation}
\left\langle H^{\prime }\right\rangle _{av}=\frac{e^{\beta \sqrt{\lambda +1}%
}+1}{4\sqrt{\lambda +1}\left( e^{\beta \sqrt{\lambda +1}}-1\right) },
\end{equation}
that already satisfy equation (\ref{eq:<H>'_b}).

\section{Conclusions}

\label{sec:conclusions}

Earlier derivations of the HFT in statistical
mechanics\cite{FC95,R07,F14b} resorted to the sum over states of
the Hamiltonian operator $H$ and, consequently, required the HFT
in quantum mechanics ($T=0$)\cite{F39}. This fact motivated an
unnecessary discussion about the validity of the mathematical
proofs in the case of degeneracy\cite{R07}. Here, on the other
hand, we have developed some of those expressions without taking
into account neither the eigenstates nor the eigenvalues of the
Hamiltonian operator thus showing that the occurrence of
degeneracy is irrelevant. It is worth noticing that the main
expressions derived by Fan and Chen\cite{FC95}, by Rai\cite{R07},
as well as present ones, apply under exactly the same conditions.
They merely differ in the strategies for their derivation. In
addition to what has just been said, we have derived a simpler
expression for the HFT in the canonical ensemble
(\ref{eq:<H>'_b}).

In the case of the variation of the statistical weight of an observable $A$
we have also derived a simple expression when $[H,A]=0$ which has proved
useful for the derivation of the HFT in the grand canonical ensemble. For
the non-commuting case our expression, although simple, depends on an
operator $F$ that cannot be obtained in closed form for the general case. It
is worth noticing that the result of Fan and Chen\cite{FC95} depends on an
operator $\hat{O}$ that exhibits the same difficulty. We have also argued
that the approach of Rai\cite{R07} is by no means a generalization of that
one of Fan and Chen\cite{FC95} that applies to degenerate states provided
that one chooses the correct eigenvectors of the Hamiltonian operator\cite
{F04}.

\section*{Acknowledgements}

The research of P.A. was supported by Sistema Nacional de Investigadores
(M\'exico).

\appendix

\numberwithin{equation}{section}

\section{Convergence of the traces}

\label{sec:appendix}

In principle, the trace in equation (\ref{eq:tr(HHH)}) may not exist in the
case of an infinite basis set $\left\{ \left| i\right\rangle ,i=1,2,\ldots
\right\} $. To overcome this problem we define
\begin{equation}
H_{M}=\sum_{i=1}^{M}\sum_{j=1}^{M}H_{ij}\left| i\right\rangle \left\langle
j\right| ,\;H_{ij}=\left\langle i\right| H\left| j\right\rangle ,
\label{eq:H_N}
\end{equation}
so that
\begin{equation}
H=\lim\limits_{M\rightarrow \infty }H_{M}.  \label{eq:H=lim_H_N}
\end{equation}
Therefore, if we repeat the argument given in section~\ref{sec:HFT
statistical} we have
\begin{equation}
\frac{\partial }{\partial \lambda }\mathrm{tr}\left[ f\left( H_{M}\right)
\right] =\mathrm{tr}\left[ H_{M}^{\prime }D_{H_{M}}f\left( H_{M}\right)
\right] ,  \label{eq:dtr(H_N)/lambda}
\end{equation}
and recover equation (\ref{eq:tr(f)'}) in the limit $M\rightarrow \infty $.

A more detailed discussion of infinite sums like (\ref{eq:H=lim_H_N}) is
available in a recent comprehensible paper\cite{BBP20} and the references
therein.

\end{document}